\documentclass{article}
\usepackage{amsthm,amsmath,amssymb}
\usepackage{mathrsfs}
\usepackage{bm}
\usepackage{geometry}
\usepackage{graphicx}
\usepackage{caption}
\usepackage[affil-it]{authblk}
\usepackage{cite}
\begin{document}
\title{An Extension of Many-Interacting-Worlds Method on Non-Guassian Model}
\author{Wen Chen}
\author{An Min Wang \thanks{anmwang@ustc.edu.cn}}
\affil{Department of Modern Physics, University of Science and Technology of China}
\maketitle
\begin{abstract}
Discussions about whether quantum theory is determinism or indeterminism has lasted for a century. A new approach to standard quantum mechanics
called many-interacting-worlds method based on many-worlds interpretation and de Broglie-Bohm mechanics provided the possibility to demonstrate
probability from
deterministic universe.The many-interacting-worlds method has been proved successful in the ground state of harmonic oscillator. In this article we
extend this method to one dimensional Coulomb potential and construct a corresponding empirical density function. We also provide a theoretical
proof of the
convergence of density function. Our numerical simulation of one dimensional Coulomb potential in the first excited state obtains the consistent
result with standard quantum mechanics and shows the applicability of many-interacting-worlds method. This research provides the possibility to
extend
many-interacting-worlds method to non-Gaussian quantum systems.
\end{abstract}
\section{Introduction}
%measurenment problem
The measurement problem is always the fundamental problem in quantum related discussion. In the history of quantum mechanics, kinds of theories
typically differ from the interpretation of the measurement process. The orthodox quantum mechanics also known as Copenhagen interpretation explained
the measurement process by the collapse of the wave function as an independent law which is considered to be unprofessional by Bell . The many-worlds
interpretation developed by Hugh Everett described the measurement by ``branching'' which had to face qualitative problem and quantitative problem
about probability \cite{saunders2010many}\cite{vaidman2014quantum}. The Bohmian mechanics as a deterministic theory viewed measurement process as the
evolution of wave function determined by Schr\"{o}dinger equation and the evolution of trajectory determined by guidence equation
\cite{pladevall2019applied}\cite{sebens2015quantum}, but controversial for ``hidden variables''. Other interpretations were mentioned in
\cite{ringbauer2017reality}\cite{dewitt2015many} and several locally deterministic interpretations were reviewed in \cite{waegell2017locally}. All
ontic and epistemic interpretations argue about what the reality of wave function is, in other words, what object the quantum state is
\cite{ringbauer2017reality}. Measurement is the key to connect the reality and wave function or probability \cite{merali2015really}. The acceptance of the measurement as a
special behavior in the evolution of quantum system like Copenhagen interpretation may bring the discussion into philosophy about ontology and
epistemology \cite{ringbauer2017reality}. We will not argue about these opinions like Einstein and Bohr but using another method to show the
connection between reality and probability. \par
Hall, Deckert, and Wiseman proposed a many-interacting-worlds (MIW) method through  a huge, but finite, number of classical ``worlds'' and quantum
effects arise from the interaction between these worlds \cite{hall2014quantum}. The worlds in MIW are definitely different from the many-worlds
interpretation since classical worlds mean determinism in every world and never branch out. If all equally weighted worlds represent reality, then
the probabilities come from the ignorance of which world the observers actually occupy. The reality or ontology in MIW method is the macroscopic
properties of worlds like world configurations and velocities described by probability density while more detailed discussion in
\cite{sebens2015quantum} called Prodigal QM which is an early idea blending many-world and Bohmian mechanics. We could say that the observers will
not specialize the measurement process in MIW since it's involved in the evolution of entire system according to Bohmian mechanics while MIW method
interpreted the evolution classic-like by introducing interaction potential. \par
The physical part of this method mainly constructed on de Broglie-Bohm mechanics but represented with respect to probability density function instead
of wave function, which made it easy to discuss the transformation from continuous density to discrete empirical density. According to the quantum potential in Bohm mechanics\cite{bohm1952suggested} and the discussion in \cite{hall2014quantum}, since we mainly work on the average of energy in the following sections, the interworld potential for $N$ countable ``worlds'' can be replaced by%unfold the theory
\begin{equation}
U_{N}(\mathbf{X})=\sum^{N}_{n=1}\sum^{K}_{k=1}\left.\frac{\hslash^{2}}{8m^{k}P(\mathbf{q})^{2}}\left(\frac{\partial P(\mathbf{q})}{\partial
q^{k}}\right)^{2}\right|_{\mathbf{q}=\mathbf{x}_{n}},
\end{equation}
where $\mathbf{X}=\{\mathbf{x}_{1},\dots ,\mathbf{x}_{N}\}$  denotes the world configurations,  $\mathbf{x}_{n}=\{x^{1}_{n}(t),\dots ,x^{K}_{n}(t)\}$
denotes the world-particle position, $m^{k}$ is the mass of the $k$th particle, $P(\mathbf{q})$ is the probability density of system. Our purpose is
to find
some proper empirical distribution
\begin{equation}
\mathbb{P}_{N}(A)=\frac{\#\{n:x_{n}\in A\}}{N},\label{empirical distribution}
\end{equation}
for any Borel set $A\subset\mathbb{R}$, whose density $P^{*}_{N}(A)$ can approximate the probability density $P(\mathbf{X})$, where $\# A$ denotes
the number of elements in the set $A$.
We expect that the quantum effects would improve as $N\to\infty$. A toy model related to ground state of harmonic oscillator was simulated using MIW
method in \cite{hall2014quantum}.
In this article we will not use dynamical MIW algorithm \cite{sturniolo2018computational}\cite{cruz2018quantum}\cite{goldstein2014quantum} to evolve
the world-particle and the classical mechanics in per ``world'' is also built on the interworld interaction potential, so we can pay little attention
on time variable $t$.\par
Mathematically, ones can interpret this method  from probability theory  since the approximation of probability is an important part in MIW method.
The intelligent idea in this method is countable classical ``worlds'' reflected in the $N$ of empirical distribution. How to find the proper
empirical density is the key point when we apply MIW method to a given system. Stein's method is a good choice for density approximation. Actually
the relation between MIW method and stein's method has been studied deeply in
\cite{mckeague2016stein}\cite{chen2020optimal}\cite{mckeague2016convergence}.\par
The higher excited states \cite{mckeague2016stein}\cite{ghadimi2018nonlocality} and higher dimensional models \cite{herrmann2017ground} of harmonic oscillator have been
simulated. All of these show the succeed of MIW method as an approach to standard quantum mechanics at least in harmonic oscillator model. \par
However all of these discussion about harmonic oscillator rely on the close relation between the solution of harmonic oscillator and standard
Gaussian \cite{herrmann2017ground}. The framework and foundation of stein's method on normal approximation is also more detailed than other
distribution. When it comes to other models such as hydrogen atom which was also a perfectly solved problem in quantum history, we need to consider
the applicability of MIW method. Mathematically saying, we need to consider the approximation of non-gaussian distribution. In this article we will
test the one-dimensional Coulomb potential model in the first excited state using MIW method. As a necessary part in approximation of density, we
also provide a proof of the convergency when the number of ``worlds'' $N\rightarrow\infty$. Nevertheless, we try to discuss the generality when apply
MIW method to different cases.\par
Before we list our results, we should explain why we focus on the one-dimension case and the first excited state of Coulomb potential model. Our
purpose is to test MIW method with hydrogen model and the results will be reflected in the approximation of empirical distribution to hydrogen atom's
probability
density. The one dimensional case is the most intuitive and mathematically simplified. Higher dimension will not only greatly increase the complexity
of analytical calculation but also be difficult to graph the comparison of simulated results. %expand to discuss
For choosing the first excited state, it's easy to derive that the ground state of one dimensional hydrogen atom is trivial \cite{loudon1959one}.
The node problem in the first excited state of one dimensional hydrogen atom case is different from non-ground states of one dimensional harmonic oscillator, since there is an infinity potential barrier at $x=0$. This constraint will turn to a special boundary condition at $x=0$ which forces us to discuss interworld potential in half in the following sections.
\section{Interworld interaction potential}
According to MIW method, the Hamiltonian of a given system is given by
\begin{equation}
H_{N}(\mathbf{X},\mathbf{P}):=\sum_{n=1}^{N}\sum_{k=1}^{K}\frac{(p_{n}^{k})^{2}}{2m^{k}}+\sum_{n=1}^{N}V\left(\mathbf{x}_{n}\right)+U_{N}(\mathbf{X}),
\end{equation}
where the first term is the kinetic energy with the definition of momentum $\mathbf{P}=\{\mathbf{p}_{1},\dots,\mathbf{p}_{N}\}$ and
$\mathbf{p}_{n}=\{p_{n}^{1},\dots,p_{n}^{K}\}$, the second term is the potential energy of given system. In our case we default $K=1$ which means
only one particle per ``world''
and totally $N$ ``worlds'' in 1d Coulomb potential of the form (Gaussian units)
\begin{equation}
V\left(x_{n}\right)=-\frac{e^2}{\vert x_{n}\vert}.
\end{equation}\par
Not like 3d hydrogen atom, it turns out more difficult to completely discuss the solutions of 1d Coulomb potential \cite{loudon1959one}. Here we
display the 1st excited state solution of the form
\begin{equation}
\Psi_{1}(x)=B_{1}xe^{-\vert x\vert},
\end{equation}
where we use dimensionless unit $me^{2}/\hslash^{2}=1$ and $B_{1}=\sqrt{2}$ is determined by normalization. In the following we will focus on the
distribution density of the form
\begin{equation}
P(x)=\vert\Psi_{1}(x)\vert^{2}=2x^{2}e^{-2\vert x\vert},
\end{equation}%discuss the solution and the choose of Exp or Laplace with generalized zero-bias
called weighted Laplace distribution, which is the target function we try to approximate with some empirical distribution. But there is a singularity
problem at $z=0$, and by easily derivation its third order derivative is discontinue which makes it not smooth enough. In \cite{pike2012stein} John
Pike had developed a framework of Laplace distribution using stein's method. We will begin with the stein's equation of Laplace distribution %given
in Lemma 2.2 of \cite{pike2012stein} ($b=1$ in our case),
\begin{equation}
g'(x)-\text{sgn}(x)g(x)=\widetilde{h}(x),\  g(0)=0,
\end{equation}
where $h\in \mathcal{H}=\{h:\mathbb{R}\to\mathbb{R}\text{ Lipschitz with }\Vert h\Vert_{\infty}\leq1\}$ and $W\sim\text{Laplace}(0,1)$,
$\widetilde{h}(x)=h(x)-\mathbb{E}[h(W)]$. The solution is given by
\begin{equation}
g_{h}(x)=\left\{
\begin{aligned}
&-e^{x}\int_{x}^{\infty}\widetilde{h}(y)e^{-y}dy\ &\textrm{$x>0$},\\
&e^{-x}\int^{x}_{-\infty}\widetilde{h}(y)e^{y}dy\ &\textrm{$x<0$}.
\end{aligned}\right.
\end{equation}\par
%\begin{equation}
%g_{h}(x)=\frac{1}{2}\left(e^{x}\int^{\infty}_{x}e^{-y}\widetilde{h}(y)dy+e^{-x}\int_{-\infty}^{x}e^{y}\widetilde{h}(y)dy\right).
%\end{equation}\par
Different from other smooth continue distribution such as standard Gaussian, the form of Laplace distribution especially the absolute value makes the
solution a piecewise function. % Actually the proof of Lemma 2.1 in [ STEIN’S METHOD AND THE LAPLACE DISTRIBUTION] is a discussion in part, which is
similar to
the process of solving 1d hydrogen atom.
Thus without loss of generality, we can treat $P(x)$ in the same way. Once we discuss (weighted) Laplace distribution in part, it will degenerate to
two independent but essentially same (weighted) Exponential distribution. So we can turn to the stein's equation of weighted Exponential distribution
$P_{\pm}(x)=2x^{2}e^{\mp 2x}$ (subscript ``$+$'' means $z>0$, ``$-$'' means $z<0$)
\begin{equation}
g'(x)+\frac{P'_{\pm}(x)}{P_{\pm}(x)}g(x)=\widetilde{h}(x).
\end{equation}\par
%and the solution is given by
%\begin{equation}
%g_{h}(x)=-\frac{1}{P_{\pm}(x)}\int_{x}^{\infty}\widetilde{h}(y)P_{\pm}(y)dy.
%\end{equation}\par
If we want to perform the zero-bias transformation based on the stein's equation to construct another distribution, we must deal with the singularity
at $z=0$ which is excluded by stein's equation but continue where actually $P_{\pm}(0)=0$. \cite{mckeague2016stein} provided a generalized zero-bias
transformations that solved the singularity in higher excited states of 1d harmonic oscillator or called two-sided Maxwell case. We will try to apply
this method to our case. Follow the Definition 3.1 in \cite{mckeague2016stein}, we say that a random variable $W^{*}$ has %if change the name of
definition
\emph{b-generalized-zero-bias distribution} of $W$ if
\begin{equation}
\sigma^{2}\mathbb{E}\left[\frac{g'(W^{*})}{b(W^{*})}\right]=\mathbb{E}\left[\frac{\text{sgn}(W)g(W)}{b(W)}\right]
\end{equation}
for $\sigma^{2}=\mathbb{E}[W^{2}/b(W)]<\infty$, where $W$ is a symmetric random variable with density $P(x)=b(x)e^{-2\vert x\vert}$ and
$b:\mathbb{R}\to\mathbb{R}$ a non-negative function. The density of $W^{*}$ is
\begin{equation}
P^{*}(z)=\frac{1}{\sigma^{2}}b(z)\mathbb{E}\left[\frac{\text{sgn}(W)}{b(W)}1_{\vert W\vert\geq \vert z\vert}\right].
\end{equation}
In fact $z\leq0$ or $z\geq0$ makes no difference when we discuss the process of construction so that we will only focus on $z\geq0$ while $z\leq0$
just need adjust the sign. The proof is similar to the Proposition 2.1 in \cite{chen2011normal}
\begin{equation}
\begin{aligned}
\sigma^{2}\int_{0}^{\infty}\frac{g'(z)}{b(z)}\frac{1}{\sigma^{2}}b(z)\int_{z}^{\infty}\frac{1}{b(w)}b(w)e^{-2w}dwdz
&=\int_{0}^{\infty}g'(z)\int_{z}^{\infty}e^{-2w}dwdz\\
&=\int_{0}^{\infty}\int_{0}^{w}g'(z)e^{-2w}dzdw\\
&=\int_{0}^{\infty}\frac{g(w)}{b(w)}b(w)e^{-2w}dw.\nonumber
\end{aligned}
\end{equation}\par
Following the Proposition 3.5 in \cite{mckeague2016stein}, we can also construct a empirical density function using the \emph{b-generalized-zero-bias
distribution} method. Given a empirical distribution $\mathbb{P}_{N}(A)$ defined in Eq.(\ref{empirical distribution}), and let
$x_{1}>x_{2}>\dots>x_{N}>0$, $x_{0}=\infty$, $x_{N+1}=0$. The corresponding density function $\mathbb{P}^{*}_{N}$ is given by
\begin{equation}
P^{*}_{N}(x)\propto b(x)\sum_{i=1}^{n}\frac{1}{b(x_{n})}
\end{equation}
for $x_{n+1}<x\leq x_{n}$ ($n=1,\dots,N-1$). \par
Next we aim at the interworld interaction potential. We begin with 1d Coulomb potential and notice that
\begin{equation}
\begin{aligned}
\left(\sum_{n=1}^{N}\frac{1}{x_{n}}\right)^{2}
&=\left[\sum_{n=1}^{N}\frac{x_{n}^{2}-x_{n+1}^{2}}{x_{n}(x_{n}^{2}-x_{n+1}^{2})}\right]^{2}\\
&=\left[\sum_{n=1}^{N}\left(\frac{1}{x_{n}(x_{n}^{2}-x_{n+1}^{2})}-\frac{1}{x_{n-1}(x_{n-1}^{2}-x_{n}^{2})}\right)x_{n}^{2}\right]^{2}\\
&\leq\left[\sum_{n=1}^{N}1^{2}\right]\left[\sum_{n=1}^{N}\left(\frac{1}{x_{n}(x_{n}^{2}-x_{n+1}^{2})}-\frac{1}{x_{n-1}(x_{n-1}^{2}-x_{n}^{2})}\right)^{2}\frac{x_{n}^{4}}{1^{2}}\right]\\
&=N\sum_{n=1}^{N}\left(\frac{1}{x_{n}(x_{n}^{2}-x_{n+1}^{2})}-\frac{1}{x_{n-1}(x_{n-1}^{2}-x_{n}^{2})}\right)^{2}x_{n}^{4}\nonumber
\end{aligned}
\end{equation}
with equality if and only if
\begin{equation}
\frac{1}{x_{n}^{2}}=\alpha\left(\frac{1}{x_{n}(x_{n}^{2}-x_{n+1}^{2})}-\frac{1}{x_{n-1}(x_{n-1}^{2}-x_{n}^{2})}\right)\label{condition1}
\end{equation}
for some constant $\alpha$. Let's rewrite $P^{*}_{N}(x_{n})$ as
\begin{equation}
P^{*}_{N}(x_{n})=\frac{\alpha}{N+1}\frac{x_{n}}{x_{n}^{2}-x_{n+1}^{2}}
\end{equation}
since $x_{N+1}$ is counted in the interaction which means we have $N+1$ worlds here.\par
Using $\frac{P^{*'}_{N}(x_{n})}{P^{*}_{N}(x_{n})}=\frac{P^{*}_{N}(x_{n-1})-P^{*}_{N}(x_{n})}{(x_{n-1}-x_{n})P^{*}_{N}(x_{n})}$, the $U_{N}(\mathbf{X})$ is approximated by
\begin{equation}
U_{N}(\mathbf{X})\approx4\sum_{n=1}^{N}\frac{\hslash^{2}}{8m}\left(\frac{1}{x_{n}(x_{n}^{2}-x_{n+1}^{2})}-\frac{1}{x_{n-1}(x_{n-1}^{2}-x_{n}^{2})}\right)^{2}x_{n}^{4}\geq\frac{1}{N}\frac{\hslash^{2}}{2m}\left(\sum_{n=1}^{N}\frac{1}{x_{n}}\right)^{2}.\label{UN}
\end{equation}\par
Then we put $U_{N}(\mathbf{X})$ and $V_{N}(\mathbf{X})$ together to display the average Hamiltonian of $N+1$ worlds at 1st excited state
\begin{equation}
\begin{aligned}
H_{N}(\mathbf{X})&=\frac{1}{N+1}\left(U_{N}(\mathbf{X})+V_{N}(\mathbf{X})\right)\\
&\geq\frac{1}{N+1}\left[\frac{1}{N}\frac{\hslash^{2}}{2m}\left(\sum_{n=1}^{N}\frac{1}{x_{n}}\right)^{2}-\sum_{n=1}^{N}\frac{e^{2}}{x_{n}}\right]\\
&=\frac{1}{N(N+1)}\frac{\hslash^{2}}{2m}\left[\sum_{n=1}^{N}\frac{1}{x_{n}}-\frac{me^{2}N}{\hslash^{2}}\right]^{2}-\frac{me^{4}N}{2\hslash^{2}(N+1)}\\
&\geq-\frac{me^{4}N}{2\hslash^{2}(N+1)}\nonumber
\end{aligned}
\end{equation}
with the second equality if and only if
\begin{equation}
\sum_{n=1}^{N}\frac{1}{x_{n}}=\frac{me^{2}N}{\hslash^{2}}=N,\label{condition2}
\end{equation}
where the second equality deu to dimensionless unit.\par
Now the minimun of $H_{N}$ is the energy of 1d hydrogen atom at 1st excited state as expected when $N\to\infty$. In next section we will show that
the conditions Eq.(\ref{condition1}) and Eq.(\ref{condition2}) can both be satisfied and will derive the recursion for simulation. In order to
determine $\alpha$,
we substitute the conditions for equality to holds into Eq.(\ref{UN}) and obtain $\alpha=1$.\par
Though the discussion above is only half of the whole system, the symmetry ensure the same process can work in $(-\infty,0)$. The total number of worlds is $2N+1$ since we fix the world $x_{N+1}$ at $x=0$ as a boundary condition.
\section{Simulation of 1d 1st excited state}
In this section we will focus on the comparison between $P^{*}_{N}(\mathbf{X})$ and $P(x)$. %Here we rewrite $P(x)$ with respect to $x$ not $z$ and
using dimensionless units for convinience which means we will set $me^{2}/\hslash=1$. Under these preconditions we have $P(x)=2x^{2}e^{-2x}$ for
$x\geq0$.
To solve the world configurations, we take the sum of both side of Eq.(\ref{condition1}) and obtain the recursion
\begin{equation}
x_{n+1}^{2}=x_{n}^{2}-\left(x_{n}\sum_{i=1}^{n}\frac{1}{x_{i}^{2}}\right)^{-1}.\label{recursion1}
\end{equation}
Using the boundary condition $x_{N+1}=0$ we can theoretically solve a complex equation with respect to $x_{1}$
\begin{equation}
x_{N}(x_{1})^{3}\left(\sum_{i=1}^{N}\frac{1}{x_{i}(x_{1})^{2}}\right)-1=0,\label{recursion2}
\end{equation}
which means we obtain all world configurations.\par

\begin{figure}[ht]
\centering
\includegraphics[width=8.6cm]{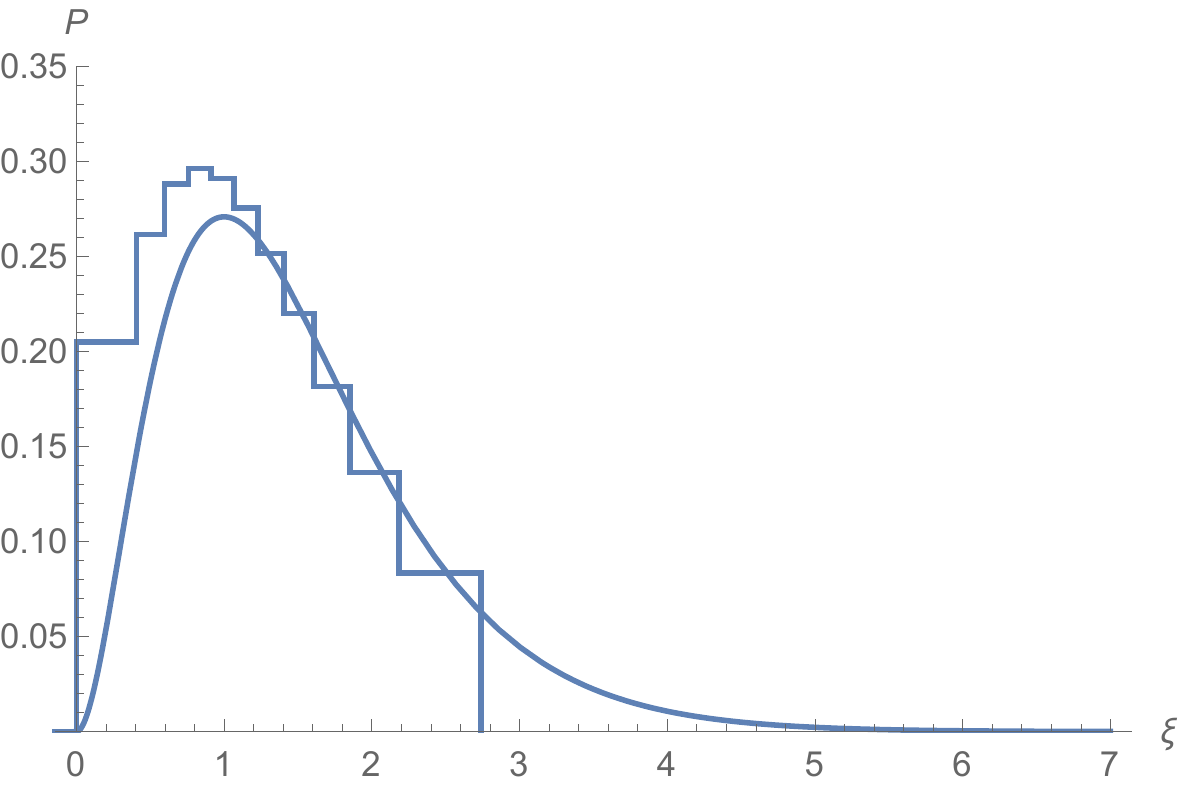}
\caption{Half of 1d Coulomb potential model for $N=11$ ($x\geq0$) at 1st excited state. The stepped curve is
$P^{*}_{N}(x_{n})=x_{n}(N+1)^{-1}(x_{n}^{2}-x_{n+1}^{2})^{-1}$ with respect to dimensionless world configurations $x_{1},\dots,x_{N}$ solved by
Eq.(\ref{recursion2}) and
$\sum_{n=1}^{N}P^{*}_{N}(x_{n})(x_{n}-x_{n+1})\approx0.54$. The smooth curve is $P(x)=2x^{2}e^{-2x}$ which is the dimensionless solution of 1d 1st
excited state hydrogen atom in $x\geq0$.}\label{fig1}
\end{figure}\par
For small values of $N$, Eq.(\ref{recursion2}) can be solved analytically as shown in Fig.\ref{fig1} when $N=11$, where $P(x)$ is plotted by smooth
curve. Though it seems that the approximation to $P(x)$ is not perfect before the local maximum when $N$ is small, the probability
$\mathbb{P}_{N,+}=\sum_{n=1}^{N}P^{*}_{N}(x_{n})(x_{n}-x_{n+1})$ is approaching to $0.5$ which is exactly $\int_{0}^{\infty}P(x)dx$ with increaing
$N$ and we will prove the convergency in next section. With the convergency, it's predictable that $x_{1}$ is increasing with $N$ which means the
area
$\int_{x_{1}}^{\infty}P(x)dx$ is decreasing and so is the difference area before local maximum.\par

\begin{figure}[ht]
\centering
\includegraphics[width=8.6cm]{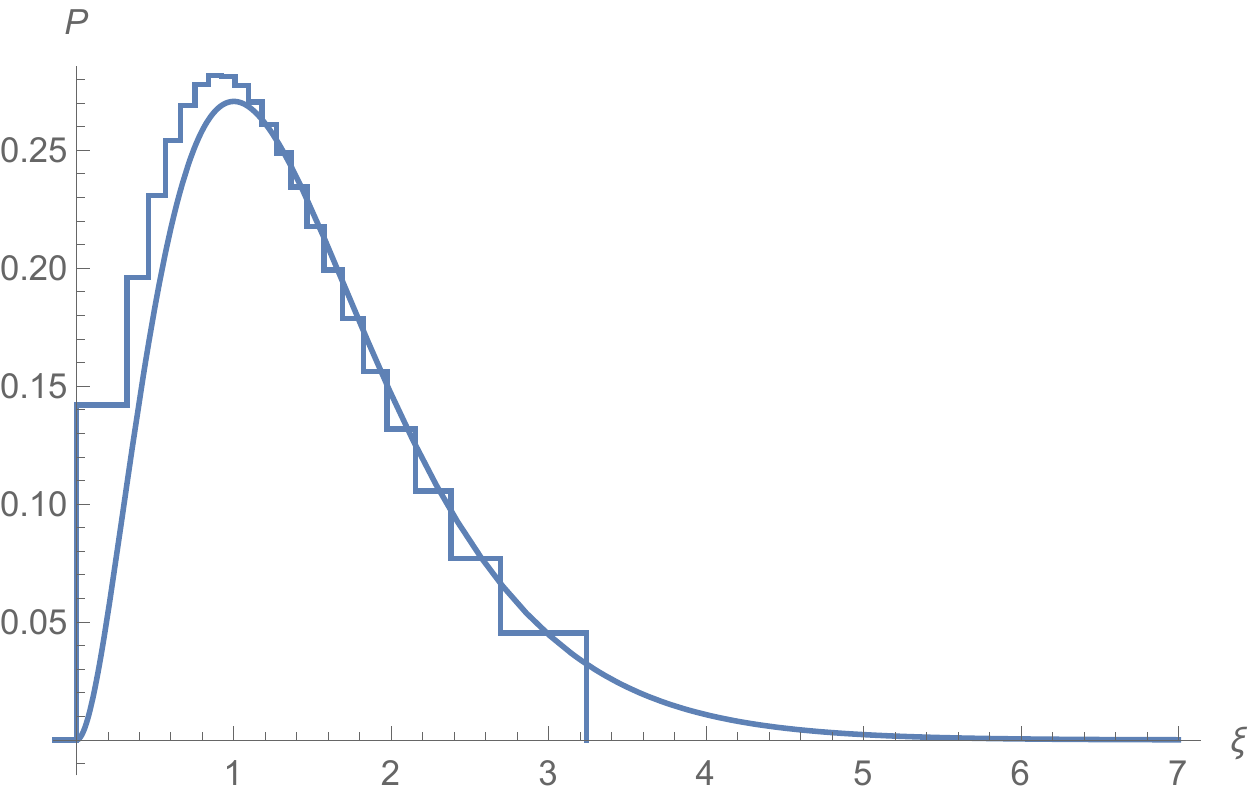}
\caption{Half of 1d Coulomb potential model for $N=21$ ($x\geq0$) at 1st excited state. The stepped curve is
$P^{*}_{N}(x_{n})=x_{n}(N+1)^{-1}(x_{n}^{2}-x_{n+1}^{2})^{-1}$ with respect to dimensionless world configurations $x_{1},\dots,x_{N}$ solved by
Eq.(\ref{recursion2}) and
$\sum_{n=1}^{N}P^{*}_{N}(x_{n})(x_{n}-x_{n+1})\approx0.526$. The smooth curve is $P(x)=2x^{2}e^{-2x}$ which is the dimensionless solution of 1d 1st
excited state hydrogen atom in $x\geq0$.}\label{fig2}
\end{figure}\par
Larger values of $N$ is shown in Fig.\ref{fig2} with numerical solution. The approximation is better as expected but it took 2 weeks to calculate on
personal computer. The symmetry of the solution ensure the results in $x<0$.
\section{Proof of convergency}
As shown in previous section, it's natural to expect that $P(x)=\lim_{N\to\infty}P^{*}_{N}(x)$. Before the proof we need to bound $x_{N}$ using
condition Eq.(\ref{condition2}) and Eq.(\ref{recursion2}).
\begin{equation}
x_{N}^{3}N=x_{N}^{3}\frac{1}{N}\left(\sum_{n=1}^{N}\frac{1}{x_{n}}\right)^{2}\leq1=x_{N}^{3}\sum_{n=1}^{N}\frac{1}{x_{n}^{2}}\leq
x_{N}^{3}\frac{1}{x_{N}}\sum_{n=1}^{N}\frac{1}{x_{n}}=x_{N}^{2}N
\end{equation}
which means
\begin{equation}
\frac{1}{\sqrt{N}}\leq x_{N}\leq\frac{1}{\sqrt[3]{N}}.
\end{equation}
The inequality become equality if and only if $N=1$. Thus we can assume that $x_{N}=O(1/\sqrt[a]{N})$ where $2<a<3$. We begin with the differential
equation
\begin{equation}
\begin{aligned}
\frac{P^{*}_{N}(x_{n})-P^{*}_{N}(x_{n+1})}{x_{n}-x_{n+1}}
&=\frac{x_{n}^{2}\displaystyle\sum_{i=1}^{n}\frac{1}{x_{i}^{2}}-x_{n+1}^{2}\sum_{i=1}^{n+1}\frac{1}{x_{i}^{2}}}{(N+1)(x_{n}-x_{n+1})}\\
&=\frac{1}{N+1}\left[\frac{1-x_{n}}{x_{n}(x_{n}-x_{n+1})}\right]\\
&=\frac{1}{N+1}\left[\frac{x_{n}+x_{n+1}-x_{n}x_{n+1}-x_{n}^{2}}{x_{n}(x_{n}^{2}-x_{n+1}^{2})}\right]\\
&=\frac{1}{N+1}\left[\frac{x_{n}-1}{x_{n}(x_{n}+x_{n+1})}\right]+2\left(\frac{1}{x_{n}}-1\right)P^{*}_{N}(x_{n}).\label{ode}
\end{aligned}
\end{equation}
Now let's discuss the first term in last equality. We focus on the function
\begin{equation}
f(x_{n})=\frac{x_{n}-1}{x_{n}(x_{n}+x_{n+1})}
\end{equation}
Since the sequence $x_{n}>0$ and $\lim_{N\to\infty}x_{N}=0$, we have $\text{max}\vert f(x_{n})\vert=\vert f(x_{N})\vert$. The order can be determined
by
\begin{equation}
\frac{1}{N+1}\left\vert\frac{x_{n}-1}{x_{n}(x_{n}+x_{n+1})}\right\vert\leq\frac{1}{N+1}\left\vert\frac{1}{x_{N}^{2}}\right\vert=O(N^{\frac{2}{a}-1})
\end{equation}
which means if we take $N\to\infty$, Eq.(\ref{ode}) tends to first-order ODE $P'(x)-2(1/x-1)P(x)=0$ with general solution $P(x)=Cx^{2}e^{-2x}$. To
determine $C$, we notice that
\begin{equation}
\sum_{n=0}^{N}P^{*}_{N}(x_{n})(x_{n}-x_{n+1})=\frac{1}{N+1}\sum_{n=0}^{N}\frac{x_{n}}{x_{n}+x_{n+1}}\geq\frac{1}{N+1}\sum_{n=0}^{N}\frac{x_{n}}{2x_{n}}=\frac{1}{2},
\end{equation}
and
\begin{equation}
\begin{aligned}
\int_{0}^{\infty}P^{*}_{N}(x)dx=&\frac{1}{3(N+1)}\sum_{n=0}^{N}\frac{1}{x_{n}(x_{n}^{2}-x_{n+1}^{2})}\left(x_{n}^{3}-x_{n+1}^{3}\right)\\
=&\frac{1}{3(N+1)}\sum_{n=0}^{N}\left[1+\frac{x_{n+1}^{2}}{x_{n}(x_{n}+x_{n+1})}\right]\leq\frac{1}{2}.
\end{aligned}
\end{equation}
Then we have
\begin{equation}
\frac{1}{4}C=\int_{0}^{\infty}P(x)dx=\lim_{N\to\infty}\sum_{n=0}^{N}P^{*}_{N}(x_{n})(x_{n}-x_{n+1})\geq\frac{1}{2},
\end{equation}
and by the bounded convergence theorem
\begin{equation}
\frac{1}{4}C=\int_{0}^{\infty}P(x)dx=\lim_{N\to\infty}\int_{0}^{\infty}P^{*}_{N}(x)dx\leq\frac{1}{2},
\end{equation}
which means $C=2$. Now we have shown that $P^{*}_{N}(x)$ with respect to solution of recursion Eq.(\ref{recursion1}) and Eq.(\ref{recursion2})
converges to $P(x)$ which is the solution of 1d Coulomb potential, but the rate of convergence remains uncertain. Using stein's method following
\cite{mckeague2016stein} will be an good approach to bound the rate of convergence which is not our main purpose in this article.
\section{Discussion}
As an extension of many-interacting-worlds method to non-Gaussian model, our work follows the similar process in Gaussian model by using stein's
method via the density approach to construct an proper empirical density without using wave function when the classical worlds is countable finite.
With some mathematical techniques we obtained the interaction potential, and also calculated the world configurations and energy described by
density. We have analytically proved the availability of MIW method on one dimensional Coulomb potential. But what we have done is the framework of
the Coulomb model which does not include the whole MIW method since the determinism of worlds need to be discussed in future work especially the
velocity and dynamic. Yet we still ignore the spin and entanglement problem (discussed in \cite{elsayed2018entangled} using Bohmian trajectories) for
the unclear reality of wave function. \par
With the application of MIW method in several potential systems, we should consider the generality of this method. Here we propose a question: if MIW
method works for any given potentials? If we simplify this question, till now the $x^{2}$ potential and $x^{-1}$ potential have been simulated, but
how about
$x^{r}$ potential when $-1<r<2$. The intuitive idea is to follow the process before. Firstly to obtain the solution of Schr\"{o}dinger equation with
$x^{r}$ potential. Then we can approximate the probability and try to construct the interaction potential to test the availability of MIW method.\par
If we follow the original idea in \cite{hall2014quantum}, it just needs \emph{a suitably smoothed version of the empirical density} to approximate
$P(x)$ and its derivatives. In our case the singularity of $P(x)$ at $x=0$ makes it not smooth but we find a suitably smoothed empirical density
$P_{N}(x)$ to approximate it piecewise. It seems remain lots of option about the empirical density but when it comes to a specific potential system
it's not easy to figure out the proper density. Especially when the form of $P(x)$ is not common with singularity or piecewise even discontinuous,
the smooth of corresponding empirical density will be hard to guarantee that means for some strange potential system MIW method may fail. Obviously
we need to consider the framework of MIW method about whether it is available in the system that can be solved by standard quantum mechanics at
least. \par
Since most research show the connection between MIW method in stationary state problem and stein's method , we can focus on the approximation of
$P(x)$ by stein's method via density approach. The Condition 13.1 in \cite{chen2011normal} provides the constraint of $P(x)$. If for the systems that
can be solved by quantum mechanics, the corresponding $P(x)$ satisfies or piecewise satisfies the condition, we may at least conclude that MIW method
works as an approach to quantum mechanics. \par
Another crucial topic is the ability independent from standard quantum mechanics to solve the quantum systems, in other words the ability to obtain
the density independent from approximation. Since no wave function is involved in MIW method and dynamic is also based on density, it seems we cannot
internally
complete this method which means it might be regarded as an approach to quantum mechanics rather than a quantum theory. Though the ontology has been
determined, the dependence of MIW on density approach makes it controversial to be a quantum theory. We still need more framework on MIW method about
the
solution of  quantum systems. In the future we look forward to find the supplement to MIW method about the wave function or probability.
%If we regard MIW method as an interpretation of approximation to given density function $P(x)$, we will realize that MIW method can not solve a
potential system independently since we can not come out an empirical density without $P(x)$ which is the solution from quantum mechanics.
\section*{Acknowledgements}
The research was supported by National Key R\&D Program of China under Grant No.2018YFB1601402-2.
\bibliography{gai}
\bibliographystyle{plain}
\end{document}